\documentclass[aps,prd,amssymb,eqsecnum,nofootinbib,floatfix, a4paper,twocolumn]{revtex4}

\usepackage{mathrsfs}
\usepackage{latexsym,amsmath,amsfonts,amssymb}
\usepackage{bm}

\usepackage{graphicx}

\usepackage{xcolor}  

\allowdisplaybreaks

\newcommand{\be}{\begin{equation}}
\newcommand{\ee}{\end{equation}}
\newcommand{\bea}{\begin{eqnarray}}
\newcommand{\eea}{\end{eqnarray}}

\newcommand{\eq}{\eqref}

\newcommand{\sch}{{Schwarzschild \,}}

\def\k{\kappa}

\def\lam{{\lambda}}
\def\pb{{\bar \pi}}
\def\lt{\tilde l}
\def\wt{\tilde w}

\def\eps{\epsilon}
\def\d{\partial}

\def\t0{\tilde{0}}

\newcommand{\bg}{\begin{gather}}
\newcommand{\eg}{\end{gather}}
\newcommand{\bseq}{\begin{subequations}}
\newcommand{\eseq}{\end{subequations}}

\begin{document}

\title{Black holes in the long-range limit of torsion bigravity}

\author{Vasilisa \surname{Nikiforova}}
 
\affiliation{Institut des Hautes Etudes Scientifiques, 
91440 Bures-sur-Yvette, France}

\date{\today}

\begin{abstract}
We continue the study of spherically symmetric black hole solutions in torsion bigravity,
a class of Einstein-Cartan-type  gravity theories involving, besides a metric, a massive propagating torsion field.
In the infinite-range limit, these theories admit asymptotically flat black hole solutions  related to
 the presence of attractive fixed points in the asymptotic radial evolution of the metric and the torsion.
We discuss these fixed points, and the way they are approached at large radii. 
Several phenomenological aspects of asymptotically flat torsion-hairy black holes are
discussed: (i) location of the light ring and of the shadow; (ii) correction to the redshift of orbiting stars; 
and (iii) modification of the periastron precession of orbiting stars. By comparing the observable properties of
torsion-hairy black holes to existing observational data on supermassive black holes 
obtained by the Event Horizon Telescope collaboration, and by the GRAVITY collaboration, we derive 
constraints on the theory parameters of torsion bigravity. 
The strongest constraint is found to come from the recent measurement of the periastron precession of the star S2
orbiting the Galactic-center massive black hole [Astron. Astrophys. \textbf{636}, L5 (2020)], and to
be  a thousand times more stringent than  solar-system gravitational tests.
\end{abstract}

\maketitle

\section{Introduction} \label{sec1}

Black Holes (BH) are one of the most remarkable predictions of Einstein's theory of General Relativity (GR). 
They epitomize GR predictions about  strong-field gravity.
It is only recently that observational data allowed one to explore in some detail the strong-field structure of 
BHs. In particular: (i) gravitational-wave data have  probed the physics of coalescing, and ringing, BH horizons
\cite{LIGOScientific:2019fpa}; (ii) very-long-baseline-interferometry has imaged the immediate neighbourhood
of the supermassive BH at the center of the Messier 87 galaxy \cite{Akiyama:2019cqa}; and (iii) the study of stars
in highly elliptical orbits around the supermassive BH at the center of our Galaxy have quantitatively checked
the strong-field structure of  BHs \cite{Abuter:2018drb,Abuter:2020dou}.

GR has, so far, been found to be compatible with all observational data related to BHs. However, in order to 
gauge the probing power of various observational windows on BH physics, it is important to be able to
compare GR predictions to the predictions of alternative theories of gravity \cite{Berti:2015itd}. Actually, there are not many
theories of gravity predicting the existence of BHs having regular horizons, and a gravitational-field structure
different from GR BHs. [Let us, however, recall, following Ref. \cite{Barausse:2008xv}, that
even in cases where BH solutions in an alternative theory coincide with GR BHs, their perturbations
 will  generally be different from the GR ones, and thereby lead to different observable predictions.]
 For examples of BHs in alternative theories of gravity, see Refs. \cite{Sotiriou:2013qea,Doneva:2017bvd,Silva:2017uqg,Brito:2013xaa,Babichev:2015xha,Enander:2015kda}.
 
 In a recent work \cite{Nikiforova:2020sac}, we have explored the existence and structure of (spherically symmetric)
 BHs in {\it torsion bigravity}. This theory generalizes (\`a la Cartan) GR by adding to the Einstein's metric field $g_{\mu \nu}$ 
 an independent affine connection ${A^{\lambda}}_{\mu \nu}$, having a non-zero torsion tensor ${T^{\lambda}}_{[\mu \nu]}$. 
 General classes of ghost-free and tachyon-free theories involving
 a propagating torsion were introduced long ago \cite{Sezgin:1979zf,Sezgin:1981xs,Hayashi:1979wj,Hayashi:1980av,Hayashi:1980ir,Hayashi:1980qp}. Torsion bigravity is a specific, minimal class of dynamical-torsion theory containing only two excitations: a massless, GR-type
 spin-2 excitation, and a massive spin-2 one. In other words, torsion bigravity has the same excitation content as (ghost-free) bimetric gravity 
 \cite{Hassan:2011zd}. Previous work (notably \cite{Nikiforova:2009qr,Damour:2019oru,Nikiforova:2020fbz}) has indicated that
 torsion bigravity seems to define a viable alternative theory of gravity, endowed with purely geometric structures,
  and having interesting physical properties.

In the present work, we complete the study of Ref. \cite{Nikiforova:2020sac} by exploring in detail the structure of the 
{\it asymptotically flat torsion-hairy} black holes, which were found to exist in the infinite-range limit of torsion bigravity.
General torsion bigravity theories contain four parameters, denoted $c_R$, $c_F$, $c_{F^2}$ and $c_{34}$ in \cite{Damour:2019oru}.
The latter parameter does not influence the spherically-symmetric sector. We are then left with three parameters that are
conveniently parametrized in terms of: (i) a dimensionful parameter $\lambda \equiv c_F+c_R = \frac1{16 \pi G_0}$,
measuring the gravitational coupling of the massless spin-2 excitation; (ii) the dimensionless parameter $\eta \equiv c_F/c_R$ 
measuring the ratio between the coupling of the massive spin-2 field and the coupling of the massless spin-2 one; and (iii)
the inverse range (or Compton wavelength) $\k \equiv \sqrt{\eta \lambda/c_{F^2}}$ of the massive spin-2 excitation.
With this notation, the Lagrangian density of torsion bigravity reads
\be \label{lag}
L=\frac{\lam}{1+\eta} R+  \frac{ \eta \lam}{1+\eta} F + \frac{ \eta \lam}{\k^2}\left( F_{(\mu \nu)} F^{(\mu \nu)} - \frac13 F^2 \right)\,.
\ee
Here, $R$ denotes the usual scalar curvature of $g_{\mu \nu}$ , while $F \equiv g^{\mu \nu} F_{\mu \nu}$, where $ F_{\mu \nu}$
denotes the Ricci tensor of the metric-preserving, but torsionfull, affine connection ${A^{\lambda}}_{\mu \nu}$. See \cite{Damour:2019oru}
for the full definitions of these objects (which conveniently involve the choice of a vierbein  ${e_i}^\mu$).

The  long-range limit, $ \k \to 0$ looks a priori singular for the Lagrangian \eq{lag}.
[As discussed in \cite{Damour:2019oru}, this formal limit might physically correspond to
the case where $\k$ is of order of the Hubble scale $H_0$.]  However, it was
proven in Ref. \cite{Nikiforova:2020fbz} that the spherically-symmetric field-equations, and solutions, of torsion bigravity admit a smooth
limit as $ \k \to 0$. This smoothness emerges when using suitable variables, and notably the variable denoted $\pb$. See Appendix \ref{appA}
for the explicit form of the (spherically-symmetric) field equations of  torsion bigravity, on which one can explicitly see the smoothness
of the long-range limit $ \k \to 0$. 

Ref. \cite{Nikiforova:2020sac} has found that, in the infinite-range limit, $ \k \to 0$, there existed {\it asymptotically flat} BHs,
having regular horizons, and endowed with an asymptotically decaying, horizon-regular torsion field ${T^{\lambda}}_{[\mu \nu]}$.
These solutions are parametrized, besides the dimensionless theory parameter\footnote{The dimensionfull gravitational-coupling
parameter $\lambda$ does not enter these vacuum solutions.}$\eta$, by the (areal) radius $r_h$ of the BH,
and by the dimensionless parameter $\pb_0$, measuring the horizon
value of the following ($\k^2$-rescaled) combination of frame components of the curvature tensor ${F^{\lambda}}_{\mu \nu \rho}$
of the torsionfull connection ${A^{\lambda}}_{\mu \nu}$:
\bea \label{pbvsF}
 \pb =\k^{-2} \left( F_{\hat 0 \hat 1  \hat 0 \hat 1}+ F_{\hat 1 \hat 2  \hat 1 \hat 2} - F_{\hat 0 \hat 2  \hat 0 \hat 2} - F_{\hat 2 \hat 3  \hat 2 \hat 3} \right)\,. 
 \eea
 Those asymptotically-flat, torsion-hairy BHs were obtained by proving the existence of fixed points, when $r \to \infty$, of the system
 of three first-order ordinary differential equations (ODEs) describing the radial evolution (from the horizon up to spatial infinity) of
 the geometric structure $g_{\mu \nu}$, ${T^{\lambda}}_{[\mu \nu]}$. In the following, we shall successively:
 (i)  discuss (in Section II) all  the fixed points of the latter  system of ODEs, and their main properties;  
 (ii) focus (in Section III) on the properties of the particular fixed point that yields asymptotically-flat  BHs;
 and finally, (iii) study (in Section IV) the phenomenology of the latter asymptotically-flat torsion-hairy BHs, and compare some of the
 physical effects they predict to the observational results on the star S2, which orbits
the supermassive BH at the center of our Galaxy \cite{Abuter:2018drb,Abuter:2020dou}.

\section{Asymptotic fixed points of the radial evolution system}

Spherically symmetric solutions of torsion bigravity are described by four variables: $\Phi(r)$, $L(r)$, $V(r)$, $W(r)$. The
first two variables describe the metric:
\be \label{ds2}
ds^2=-e^{2\Phi}dt^2 + L^{2}dr^2 + r^2\left( d\theta^2+\sin^2\theta\, d\phi^2 \right) \,;
\ee
while the last two represent the independent components of the torsionful connection ${A^{\lambda}}_{\mu \nu}$
in the polar-type frame defined by Eq. \eq{ds2}. [See Section III of \cite{Nikiforova:2020sac} for details.]
It is then convenient to further introduce the auxiliary variables, $F(r) \equiv \Phi'(r)$ (where the prime denotes
a radial derivative), and, $\pb(r)$,
defined by Eq. \eq{pbvsF}, which means that  $\pb$ is linear in the radial derivatives of $V(r)$ and $W(r)$
(see Eq. (4.3) of \cite{Nikiforova:2020sac}). It was shown in Refs. \cite{Nikiforova:2020sac,Damour:2019oru,Nikiforova:2020fbz} 
that the five variables  $F(r) $, $L(r)$, $V(r)$, $W(r)$, $\pb(r)$ then satisfy a system of ODEs which can be reduced to:
(i) two algebraic equations that can be solved to express $F$ and $V$ in terms of $L(r)$, $W(r)$, and $\pb(r)$;
and (ii) a system of three first-order ODEs describing the radial evolution of the three variables ($L(r)$, $W(r)$, $\pb(r)$).
When discussing BH solutions it is convenient to replace the three variables ($L(r)$, $W(r)$, $\pb(r)$) by the equivalent set 
\be \label{defX}
 (X^1, X^2, X^3) \equiv (\lt ,\wt, \pb)\,.
 \ee
 Here $\lt$ and $\wt$ are defined as
 \be \label{deflw}
{\tilde l}(r)\equiv \frac{L(r)}{L_S(r)}\, ; \,  {\tilde w}(r)\equiv \frac{W(r)}{W_S(r)}\, ,
\ee
where $L_S$ and $W_S$ are the values of $L$ and $W$ for a Schwarzschild BH. We recall (see \cite{Nikiforova:2020sac})
that the geometric structure ($g_{\mu \nu}, {A^{\lambda}}_{\mu \nu}$) of a Schwarzschild BH is described by the variables
\bea \label{schw}
\Phi_S(r) &=&  +\frac12 \ln \left(1 - \frac{r_h}{r}\right)\,, \nonumber\\
F_S(r) &=& +\frac12\frac{r_h}{r(r-r_h)} \,,\nonumber\\
L_S(r) &=& \left( 1-\frac{r_h}{r} \right)^{-1/2} \,,\nonumber\\
V_S(r)  &=& \frac{ F_S(r)}{L_S(r)} =\frac12 \frac{r_h}{r^2}\left( 1-\frac{r_h}{r} \right)^{-1/2}\,, \nonumber\\
W_S(r) &=& - \frac{1}{r L_S(r)}= -\frac{1}{r} \left( 1-\frac{r_h}{r} \right)^{1/2} \,,\nonumber\\
\pb_S(r)&=&- 3 \frac{r_h}{\k^2 r^3} \,.
\eea
The three variables $X^i$, $i=1,2,3$, Eq. \eq{defX}, satisfy a first-order radial evolution system, say,
\be \label{DXsystem}
 \frac{dX^i}{dr}=F^i(r,X^j) .
 \ee
The explicit form of this system can be read off  by taking the limit $\k \to 0$ in Eqs. \eq{plwsystem} of Appendix \ref{appA}.

The large-$r $ asymptotics of the latter system yields  a Fuchsian-type system, of the form
\be \label{Fuchsk-1}
\frac{d}{dr}X^i= \frac1{r} V^i(X^j) + O\left(\frac1{r^2}\right).
\ee
Neglecting the $O\left(\frac1{r^2}\right)$ subleading correction, and introducing $\rho \equiv \ln r$ leads to
a vectorial flow equation for the $\rho$-evolution of the point ${\bf X}$ in a 3-dimensional space:
\be \label{Fuchsk0}
\frac{d}{d\rho}X^i=V^i(X^j) \,.
\ee 
Let us consider the solutions $\bf X_\infty$ of the three (algebraic) equations
\be \label{EqV0}
V^i(X^j_\infty)=0 \;. 
\ee
These solutions define the {\it fixed points} of the vectorial flow \eqref{Fuchsk0}: 
if the system \eqref{Fuchsk0} crosses a state corresponding to any of its fixed points, it will stay there forever 
\be
\frac{d}{d\rho}X^i_\infty=V^i(X^j_\infty)=0 \,.
\ee
Yet this does not tell anything about the {\it accessibility} of the fixed points. The first issue to understand, for knowing 
whether the system can reach a certain fixed point, is to assess  whether this fixed point is an attractor or not.
 Let us recall the relevant mathematics. We consider a small perturbation $\xi^i(\rho)$ around a fixed point
\be
X^i(\rho)=X^i_\infty+\eps \xi^i(\rho) \;.
\ee
Substituting the latter expression in \eqref{Fuchsk0}, and making a series expansion of the right-hand side up to the linear order, one obtains the following equation, describing linear perturbations around any fixed point
\be \label{xievol}
\frac{d}{d \rho} \xi^{i }_{\rm lin}(\rho)=\left[\frac{\d V^i}{\d X^j}\right]\left(\bf X_\infty \right)\xi^{j }_{\rm lin}(\rho)=J^i_{\;j}\left(\bf X_\infty \right)\xi^{j }_{\rm lin}(\rho) \;.  
\ee
Here 
\be
J^i_{\;j}\left(\bf X_\infty \right) \equiv \left[\frac{\d V^i}{\d X^j}\right]\left(\bf X_\infty \right)\,,
\ee
denotes the Jacobian matrix of the system \eqref{Fuchsk0}; it is made of the partial derivatives of the functions $V^i$  with respect to $X^j=\{\lt, \wt, \pb\}$  taken at the fixed point ${\bf X}_\infty=\{\lt_\infty, \wt_\infty, \pb_\infty\}$. 

To obtain the law of evolution of the linear perturbations, one introduces  the eigenvectors $E^i_{(n)}$, $n=1,2,3$, of the matrix $J^i_{\;j}$, say
\be
J^i_{\;j}E^j_{(n)}=\lambda_{(n)}E^i_{(n)} \,,
\ee
and  decomposes $\xi^{i }_{\rm lin}$ along the vector basis of eigenvectors, say:
\be
\xi^{i }_{\rm lin}(\rho)=\sum_{n}\delta_{(n)}(\rho)E^i_{(n)} \label{xidec} \;,
\ee
where it should be noted that the eigenvectors $E^i_{(n)}$ do not depend on $\rho$.
Inserting Eq. \eqref{xidec} into Eq. \eqref{xievol}, one obtains the $\rho$-evolution law for  the coefficients $\delta_{(n)}(\rho)$:
\be
\frac{d}{d \rho} {\delta}_{(n)}(\rho)=\lambda_{{n}} \,\delta_{(n)}(\rho)\,.
\ee
This gives the following equation describing the evolution of general linear perturbations $\xi^{i }_{\rm lin}$ around
a fixed point:
\be
\xi^{i }_{\rm lin}(\rho)=\sum_n c_{(n)}e^{\lam_{(n)}\rho} \,E^i_{(n)}=\sum_n C_{(n)}^i e^{\lam_{(n)}\rho} \;, \label{evallawXi}
\ee
where $c_{(n)}$ and $C^i_{(n)}$ are some constants, depending on the initial values for $\xi^{i }_{\rm lin}(\rho)$.

The first conclusion is that the attractive or repulsive character of a fixed point is determined by the eigenvalues of Jacobian matrix: 
a fixed point is an attractor (and all the trajectories from a small neighborhood of the fixed point
will end up at the fixed point) if the real parts of all the eigenvalues $\lam_{(n)}$ are {\it negative}.

 In addition, we must remember that our exact radial evolution system, Eq. \eq{DXsystem}, is to be integrated with restricted initial conditions 
 on the horizon, $r=r_h$. Indeed, Ref. \cite{Nikiforova:2020sac} has shown that a single  shooting parameter (namely $\pb_0$) can
 be specified on the horizon. Then the following radial evolution (from $r=r_h$ up to $r = +\infty$) of the three variables $X^i=\{\lt, \wt, \pb\}$
 will depend both on $\pb_0$,  on $r_h$, and on the theory parameter $\eta$. [As in Ref. \cite{Nikiforova:2020sac},
 we often use,  for convenience, units where $r_h=1$.] Therefore, besides analytically studying the attractor nature of the fixed points,
 we shall need numerical simulations to check whether the one-parameter family of  horizon initial values can belong to the basin of
 attraction of any given attractive fixed point.

\subsection{List of asymptotic fixed points of the radial evolution}
Let us now go to practice and analyze all the fixed points of the system \eqref{EqV0}. This algebraic system gives ten fixed points. Five of them are characterized by $\lt_\infty<0$ and thus are non-physical, so we will not consider them. We must indeed remember
at this stage that the physical meaning of the variable $X^1 =\lt$, defined in Eq. \eq{deflw}, is such that we need to impose $\lt(r)>0$.
We will discuss below the further conditions characterizing  {\it asymptotically flat} BH solutions. 

Among the remaining five points, one is only partially attractive: the fixed point $\left( \lt_\infty=1,\, \wt_\infty=1,\, \pb_\infty=0 \right)$  
has eigenvalues $\lam_{(n)}=\left( -1,\,-3,\,2 \right)$, so that it is repulsive in one eigen-direction.

Let us consider more in detail the four remaining fixed points. 
\begin{itemize}
\item The most interesting is the fixed point 
\be \label{Goldenfp}
\left( \lt_\infty=1,\, \wt_\infty=-1,\, \pb_\infty=\frac{6}{1+\eta} \right) \,,
\ee
describing the {\it asymptotically flat torsion-hairy}\footnote{Note that the condition $ \lt_\infty=1$ is necessary for
the asymptotic flatness of the metric.} black hole solution which was discussed in the previous paper \cite{Nikiforova:2020sac} 
(see Section X there ). The corresponding eigenvalues are 
\be
\lam_{(n)} = \left\{ -1,-\frac{1}{2} + i \frac{\sqrt{7}}{2} ,-\frac{1}{2} - i \frac{\sqrt{7}}{2} \right\}\,. \label{eigenval}  
\ee
The real parts of the eigenvalues displayed in Eq. \eqref{eigenval} are all strictly negative. 
This shows that the fixed point \eqref{Goldenfp} is a (local) attractor. 
\end{itemize}

The analytical expressions of  $\, \lt_\infty,\; \wt_\infty, \; \pb_\infty$  for the three remaining fixed points 
are rather complicated, so we will use a numerical description. They all correspond to {\it non-asymptotically-flat torsion-hairy black hole solutions}.
\begin{itemize}
\item The second fixed point {\it exists only for $\eta>1/3$}, because the values of $\eta<1/3$  give imaginary values for $\lt_\infty,\; \wt_\infty, \; \pb_\infty$. For all $\eta>1/3$, it is an attractive fixed point. The latter fact was checked numerically by computing the eigenvalues.  When $\eta$ varies from $1/3$ to infinity, the value of $\lt_\infty$ respectively varies in the interval
\be
\sqrt{2}<\lt_\infty<\sqrt{3} \;.
\ee
The fact that $\lt_\infty >1$ means that the corresponding BH is non-asymptotically-flat. Fig.\ref{fig1} shows the spectrum of values of $\lt_\infty,\; \wt_\infty, \; \pb_\infty$ as a function of the theory parameter $\eta$. 
\begin{figure}
\includegraphics[scale=0.5]{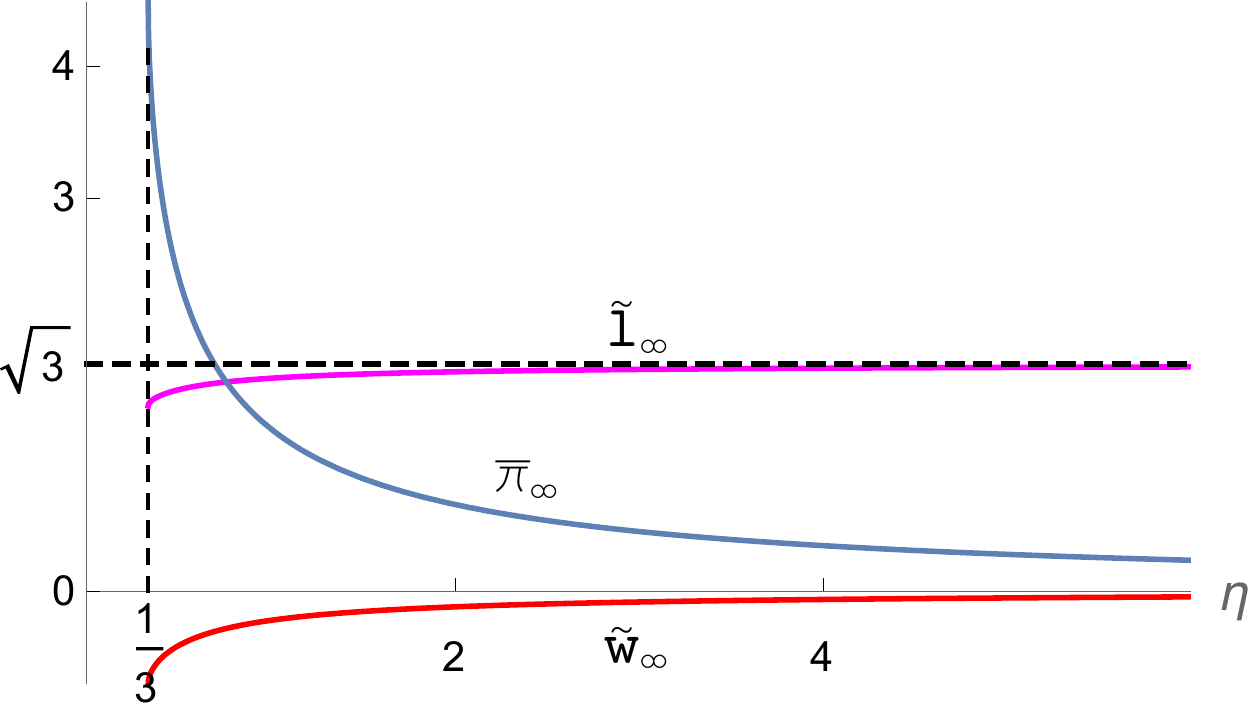}
\caption{\label{fig1}
Fixed point values $\lt_\infty,\; \wt_\infty$, and $\pb_\infty$ for the second fixed point. The value of $\wt_\infty$ varies between   $\wt_\infty(\eta=\frac13)=-1/\sqrt{2}$ and $\wt_\infty(\eta\to\infty)=0$. The value of $\pb_\infty$ varies between   $\pb_\infty(\eta=\frac13)= \frac92$ and $\pb_\infty(\eta\to +\infty)=0$. 
}
\end{figure}
\item The third fixed point also {\it makes sense only for $\eta>1/3$}, and it is an attractive fixed point for all $\eta>1/3$. When $\eta$ varies from $1/3$ to infinity, the value of $\lt_\infty$ decreases from $\lt_\infty(\eta=1/3)=\sqrt{2}$ to $\lt_\infty(\eta\to +\infty)=1.3$.
Fig.\ref{fig2} shows the spectrum of values of $\lt_\infty,\; \wt_\infty, \; \pb_\infty$ as a  function of $\eta$. 
\begin{figure}
\includegraphics[scale=0.5]{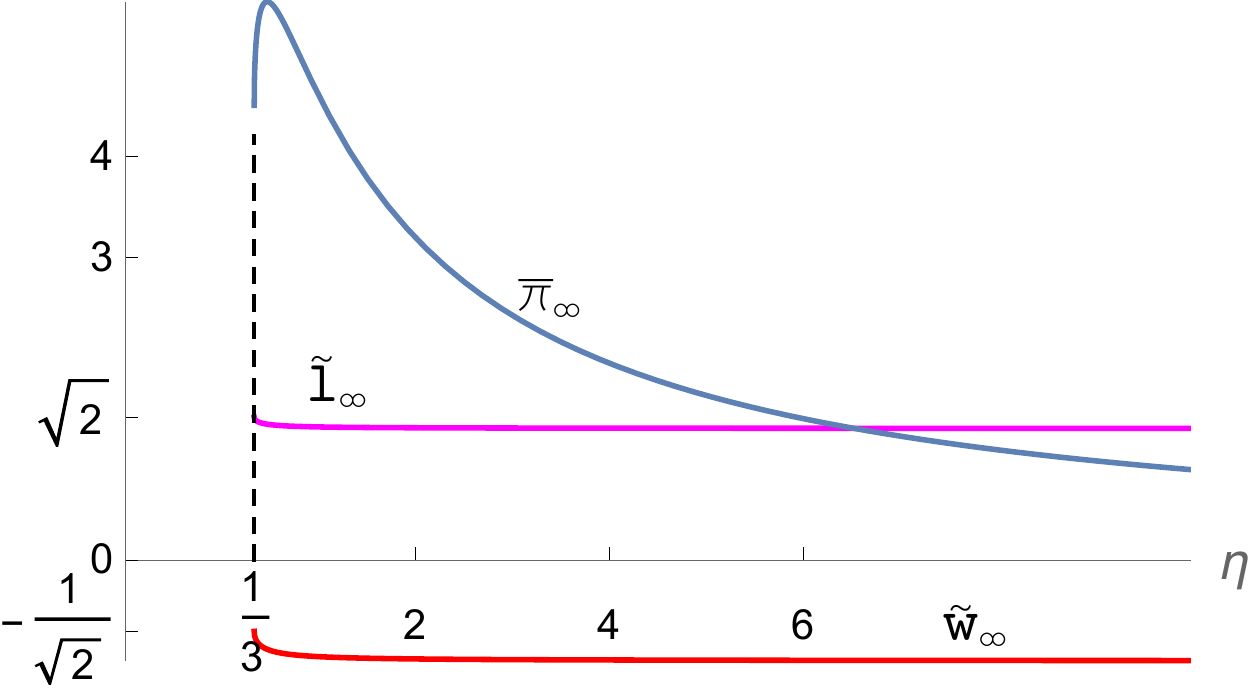}
\caption{\label{fig2}
Fixed point values $\lt_\infty,\; \wt_\infty$, and $\pb_\infty$ for the third fixed point. The value of $\wt_\infty$ varies between   $\wt_\infty(\eta=\frac13)=-1/\sqrt{2}$ and $\wt_\infty(\eta\to +\infty)=-1$. The value of $\pb_\infty$ varies between $\pb_\infty(\eta=\frac13)=\frac92$ and  $\pb_\infty(\eta\to +\infty)=0$.
}
\end{figure}
\item The last, fourth fixed point makes sense for all $\eta$'s, and it is attractive. The value of $\lt_\infty$ varies from $\lt_\infty(\eta\to0)\to+\infty$ to $\lt_\infty(\eta\to +\infty)=2.3$.
See Fig. \ref{fig3} for more details.
\begin{figure}
\includegraphics[scale=0.5]{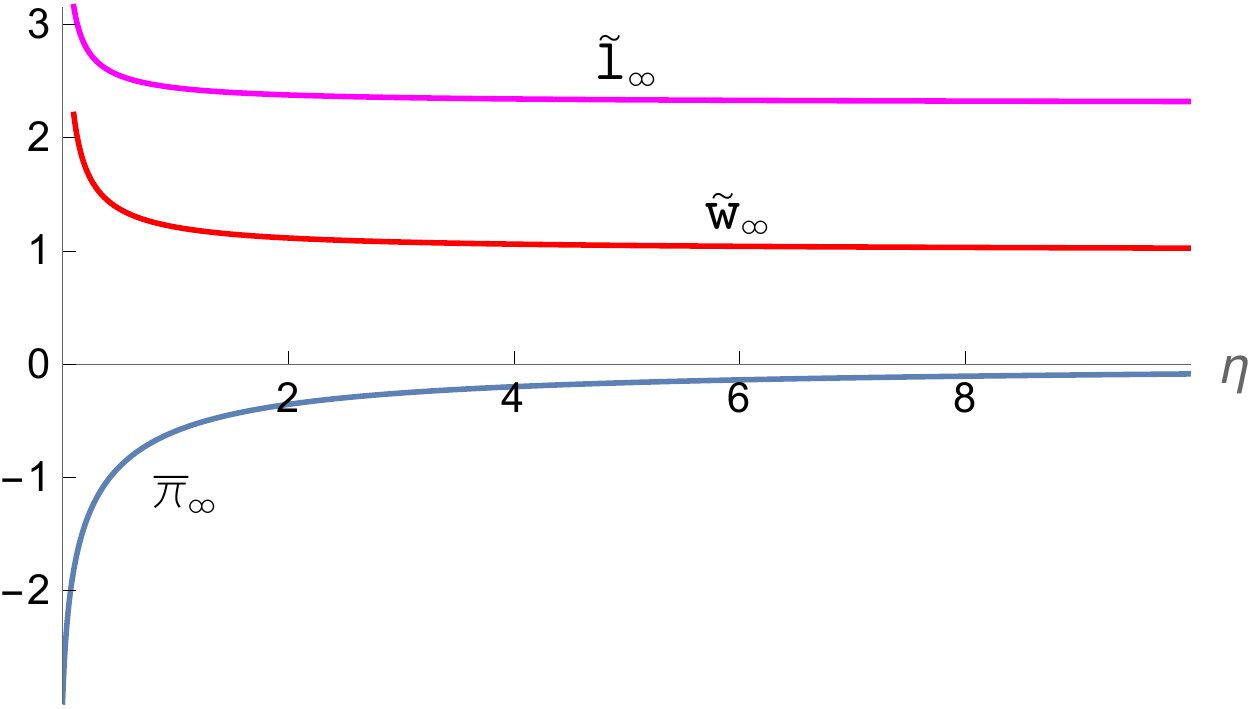}
\caption{\label{fig3}
Fixed point values $\lt_\infty,\; \wt_\infty$, and $\pb_\infty$ for the fourth fixed point. The value of $\wt_\infty$varies  between   $\wt_\infty(\eta\to 0)\to +\infty$ and $\wt_\infty(\eta\to\infty)=1$.  The value of $\pb_\infty$ varies between   $\pb_\infty(\eta\to0)\to+\infty$ and $\pb_\infty(\eta\to +\infty)=0$.
}
\end{figure}
\end{itemize}

\section{Asymptotically flat torsion-hairy BHs in the large-range limit }

Let us now analyze in some detail the asymptotically flat torsion-hairy BHs obtained with the first fixed point, i.e.:
\be
{\bf X}_{\infty} =  \left(\lt_{\infty}, \wt_{\infty}, \pb_{\infty}\right)= \left(1,-1,  \frac{6}{ \eta+1} \right)\,. \label{GoldenX}
\ee
As already mentioned, the approach towards this fixed point is characterized by the three $\eta$-independent eigenvalues 
\be
\lam_{(n)} = \left\{ -1,-\frac{1}{2} + i \frac{\sqrt{7}}{2} ,-\frac{1}{2}-i \frac{\sqrt{7}}{2} \right\}  \;.
\ee
Inserting \eqref{eigenval} into \eqref{evallawXi}, and taking into account that we are considering real solutions of the real perturbation equation
\eq{xievol}, we get, for a general real asymptotic solution, the form
\be
\xi^{i}_{\rm lin}(\rho)= C^i_{(1)} e^{-\rho}+ C^i_{(2)}e^{-\frac{\rho}{2}}e^{i\frac{\sqrt{7}}{2}\rho} + \left[  C^i_{(2)}e^{-\frac{\rho}{2}}e^{-i\frac{\sqrt{7}}{2}\rho}   \right]^\ast \,,
\ee
where $ C^i_{(1)}$ is real, and where $\ast$ means complex conjugation. This gives, in more explicit form,
\be \label{xilin}
\xi^{i }_{\rm lin}(\rho)=  2\left|C^i_{(2)}\right|e^{-\frac{\rho}{2}}\cos{\left(  \frac{\sqrt{7}}{2}\rho + \Phi^i \right)}+C^i_{(1)}e^{-\rho}  \;, 
\ee
where $\Phi^i$ are some phases. In terms of $r$ (remembering that $\rho \equiv \ln r$), this looks as follows
\be
\xi^{i }_{\rm lin}(r)=   \frac{2\left|C^i_{(2)}\right|}{r^{1/2}}\cos{\left(  \frac{\sqrt{7}}{2}\ln{r} + \Phi^i \right)} +  \frac{C^i_{(1)}}{r}  \;. \label{xilinr}
\ee
At this point, we should note that the perturbations $\xi^{i}_{\rm lin}$ were obtained in the linearized approximation. In other words, Eq. \eqref{xilinr} neglected the contributions coming from terms quadratic (and higher) in $\xi$ in the perturbation equation.
Instead of solving Eq. \eqref{xievol}, we should have solved a nonlinear equation of the form
\be
\frac{d}{d\rho}\xi^i_{\rm nonlin}=J^i_{\;j}\xi^j_{\rm nonlin} +J^i_{\;j\,k}\xi^{j}_{\rm nonlin} \xi^{k}_{\rm nonlin} + \cdots \;. \label{eqnonlin}
\ee
Let us sketch the structure of the corrections to the linearized solution  \eq{xilinr} when considering only the quadratically
nonlinear terms, and treating them  perturbatively, i.e., solving the equation
\be
\frac{d}{d\rho}\xi^i_{\rm nonlin}=J^i_{\;j}\xi^j_{\rm nonlin} +J^i_{\;j\,k}\xi^{j}_{\rm lin} \xi^{k}_{\rm lin}  \;,  \label{eqQuad}
\ee
where $\xi^{i}_{\rm lin}$ is given by \eqref{xilin}. The structure of the last, quadratic term in Eq. \eqref{eqQuad} is
\bea
&&J^i_{\;j\,k}\xi^{j}_{\rm lin} \xi^{k}_{\rm lin} = J^i_{\;j\,k}\left(\sum_n C_{(n)}^j e^{\lam_{(n)}\rho}\right) \left(\sum_m C_{(m)}^k e^{\lam_{(m)}\rho} \right)\nonumber \\
&&=\tilde{C}^i_1e^{-\rho} +\tilde{C}^i_{2+}e^{(-1+i\sqrt{7})\rho}+ \tilde{C}^i_{2-}e^{(-1-i\sqrt{7})\rho} \nonumber \\
&& +  \tilde{C}^i_{3+}e^{(-\frac{3}{2}+i\frac{\sqrt{7}}{2})\rho} + \tilde{C}^i_{3-}e^{(-\frac{3}{2}-i\frac{\sqrt{7}}{2})\rho} + \tilde{C}^i_4e^{-2\rho} \,.\label{QuadTerm}
\eea
Among the contributions contained in this quadratically nonlinear ``source term", the term 
\be
\tilde{C}^i_1e^{-\rho} \nonumber 
\ee
coincides with a fundamental solution of the homogeneous equation \eqref{xievol} with $\lam_{(n)}=\lam_{(1)}=-1$. As a consequence, 
such a resonant source term will generate a solution of \eqref{eqQuad}  of the type $e^{-\rho}\rho$. The other, nonresonant, terms in \eqref{QuadTerm} will generate exponential solutions of the same form as they are: the terms $e^{(-1\pm i\sqrt{7})\rho}$ after taking the real part will generate in the solution a term of the type $e^{-\rho}\cos{(\sqrt{7} \rho+{\rm phase})}$, while the terms $e^{(-\frac{3}{2}\pm i\frac{\sqrt{7}}{2})\rho}$ will generate a term of the type  $e^{-3\rho/2}\cos{(\frac{\sqrt{7}}{2}  \rho+{\rm phase})}$.  Finally, the quadratically nonlinear solution for $\xi^i$ will have the following structure:
\bea
\xi^{i }(\rho)&=& C^i_{(\frac{1}{2}\rm cos)}e^{-\frac{\rho}{2}}\cos{\left(\frac{\sqrt{7}}{2}\rho+\Phi^i_{(\frac{1}{2})} \right)}  \nonumber \\
&&+ C^i_{(1)}e^{-\rho}+C^i_{\rm (ln)}e^{-\rho}\rho \nonumber \\
&& + C^i_{\rm (cos)}e^{-\rho}\cos{\left(\sqrt{7}\rho+\Phi^i_{(1)}\right)} \nonumber \\
&&+ O(e^{-\frac{3\rho}{2}}) \;,
\eea
or, in terms of $r$,
\bea
\xi^{i }(r)&=& \frac{C^i_{(\frac{1}{2}\rm cos)}}{r^{1/2}}\cos{\left(  \frac{\sqrt{7}}{2}\ln{r} + \Phi^i_{(\frac{1}{2})} \right)} \nonumber \\
&& +  \frac{C^i_{(1)}}{r}+\frac{C^i_{(\rm ln)}\ln{r}}{r}  \nonumber \\
&& + \frac{C^i_{\rm (cos)}}{r}\cos{\left(\sqrt{7}\ln{r}+\Phi^i_{(1)}\right)} \nonumber \\
&&+ O(r^{-3/2}) \;. 
\label{xinonlin}
\eea
This gives the following expression for the large-$r$ behavior of ${\bf X} =  (\lt, \wt, \pb)$ 
\bea
X^i(r)&=&X^i_\infty + \frac{C^i_{(\frac{1}{2}\rm cos)}}{r^{1/2}}\cos{\left(  \frac{\sqrt{7}}{2}\ln{r} + \Phi^i_{(\frac{1}{2})} \right)} \nonumber \\
&& +  \frac{C^i_{(1)}}{r}+\frac{C^i_{(\rm ln)}\ln{r}}{r}  \nonumber \\
&& + \frac{C^i_{\rm (cos)}}{r}\cos{\left(\sqrt{7}\ln{r}+\Phi^i_{(1)}\right)} \nonumber \\
&&+ O(r^{-3/2}) \;.   \label{Xlaw} 
\eea
The last expression gives, for example, the following behavior of $\sqrt{g_{rr}} = L=  \lt/\sqrt{1-r_h/r}$ near infinity
\bea
\sqrt{g_{rr}}&=&1+\frac{C^l_{(\frac{1}{2}\rm cos)}}{r^{1/2}}\cos{\left(  \frac{\sqrt{7}}{2}\ln{r} + \Phi^l_{(\frac{1}{2})} \right)} \nonumber \\
&& +  \frac{C^l_{(1)}}{r}+\frac{C^l_{(\rm ln)}\ln{r}}{r} \nonumber \\
&&+ \frac{C^l_{\rm (cos)}}{r}\cos{\left(\sqrt{7}\ln{r}+\Phi^l_{(1)}\right)} \nonumber \\
&& + O(r^{-3/2}) \label{grr} \;.
\eea
When comparing this asymptotic behavior to the usual \sch one, namely
\be
\sqrt{g_{rr}^S}=\frac{1}{\sqrt{1-r_h/r}}=1+\frac{r_h}{2r}+O(r^{-2})\,,
\ee
the most striking difference is the fact that  the leading term in the large-$r$ expansion of $\sqrt{g_{rr}}$ is not decaying
as $O(r^{-1})$, but decays in the slower $O(r^{-1/2})$ manner, and moreover presents
 oscillations on the logarithmic scale $\ln{r}$, with a universal frequency $ \frac{\sqrt{7}}{2}$. 
By computing the corresponding solution for the potential $\Phi(r) = \frac12 \ln | g_{00}|$ describing the time-time component of the metric in torsion bigravity,
one similarly finds that the leading term in the large-$r$ expansion of $\Phi(r)$ also contains an oscillating,
slowly-decaying piece $\propto r^{-1/2} \cos{\left(  \frac{\sqrt{7}}{2}\ln{r} + {\rm phase}\right)}$.
The physical consequences of such slowly-decaying (and oscillating) contributions are discussed in the following subsections.

\subsection{Expansion of the BH solution in powers of $\eta$}

Let us show that, in the formal limit $\eta \to 0$, the metric structure (described by the functions $L(r)$ and $F(r)= \Phi'(r)$)
of asymptotically-flat BHs tend to the \sch one $L=L_S$ and $F=F_S$ (defined in Eqs. \eq{schw}).

To this end, we first remark that Eq. \eqref{Dlt} becomes very simple in the limit $\eta=0$ (with $\k=0$). One finds that it takes the following form
\be
\lt'=\frac{\lt (1-\lt^2)}{2(r-1)} \;. \label{Eqleta0}
\ee
We should solve this equation taking into account the boundary conditions at the horizon which were discussed in \cite{Nikiforova:2020sac}. Looking at the eqs. (7.8) -- (7.10) of \cite{Nikiforova:2020sac} one can see that, in the limit $\eta \to 0$, the horizon value $\lt(r=r_h) \to 1$,
independently of the value chosen for the free shooting parameter $\pb_0$.  
Since the right-hand side of \eqref{Eqleta0} equals to zero when $\lt=1$, the unique solution which satisfies the boundary conditions is
found to be
\be \label{lteq1}
\lt(r) \equiv 1 \; ({\rm when}\; \eta \to 0) .
\ee
Inserting this result (together with $\k =0$ and $\eta=0$) in Eqs. \eqref{Dwt} and \eqref{Dpb}, one then finds
that the radial behavior of $\wt$ and $\pb$ is described by the following system of two first-order equations:
\bea \label{eta0system}
r \wt'&=&-\frac{\pb^2 + 3 \pb (2 - r + r \wt) - 
  9 (2 - 2 r + ( 2 r-3) \wt)}{3( r-1)  (3 + \pb)} \;, \nonumber \\
 r  \pb'&=&3 - 2 \pb - 3 \wt - \pb \wt \,.
\eea
The large-$r$ limit of this system is a two-dimensional Fuchsian system which admits, as fixed point,
the values $(\wt_\infty,\pb_\infty)=\left( -1, 6 \right)$. These values are the  two-dimensional projection
of the $\eta=0$ limit of the fixed point \eqref{GoldenX}. The corresponding two-dimensional Jacobian matrix is
found to have the eigenvalues
\be
\lam_{(1,2)} = -\frac{1}{2} \pm i \frac{\sqrt{7}}{2}  \;. \label{eigenval2by2}
\ee
Similarly to the previous section, we can conclude from this the following law of asymptotic approach to the fixed point:
\bea
\wt_{\eta0}(r)&=&-1+\frac{A_w}{r^{1/2}}\cos{\left( \frac{\sqrt{7}}{2}\ln{r}+\Phi_w \right) }  \;, \label{wtEta0} \\
\pb_{\eta0}(r)&=&6+\frac{A_\pi}{r^{1/2}}\cos{\left( \frac{\sqrt{7}}{2}\ln{r}+\Phi_\pi \right) } \,, \label{pbEta0}
\eea
plus some terms of higher orders in $1/r$. Here, the constants $A_{w}$, $A_{\pi}$ depend on the integration constants of the
system \eq{eta0system} .

Finally, putting $\eta=0$, $\k=0$ and inserting Eqs. \eqref{lteq1}, \eqref{wtEta0}, \eqref{pbEta0} into the
algebraic equation \eq{Fsol} giving $F$ in terms of $\lt,,\wt, \pb$, one finds that 
\be
F|_{\eta=0, \, \k=0}=F_S, 
\ee
where $F_S=\frac{1}{2 ( r-1) r}$ is the \sch value. 

Having determined the structure of the solution in the limit $\eta \to 0$, we can now discuss the structure of
the solutions when $\eta$ is non-zero, but small. Let us indeed recall that Ref. \cite{Damour:2019oru} has studied
the phenomenological constraints on the torsion bigravity theory parameter $\eta$ in the limit $\k \to 0$, and has
concluded that solar-system gravity tests constraint $\eta$ to be small; more precisely (see Eq. (10.18) in \cite{Damour:2019oru})
\be \label{solarsystem}
\eta \lesssim 10^{-5} \; ({\rm from \; solar \;system \; gravitational \; tests})\,.
\ee 
It is then appropriate to work only linearly in $\eta$. We then conclude from the result 
\eqref{lteq1} that
\be \label{ltexp}
\lt(r) \equiv 1 +\eta \lt_1(r)  +O(\eta^2) \;.
\ee
One then gets a  differential equation for $\lt_1(r)$ by inserting the expansion \eq{ltexp} in \eqref{Dlt}, working at
linear order in $\eta$. Together with eq. \eqref{Dlt}, this gives 
\be
\lt'_1=-\frac{\lt_1}{r-1}+ f_{l}( \wt_{\eta0}, \, \pb_{\eta0}) + O(\eta) \;,  \label{ltCorr}
\ee
where $f_l$ is a polynomial function of its arguments (and a rational function of $r$).
Replacing $\wt_{\eta0}$ and $\pb_{\eta0}$ by solutions of the system \eq{eta0system}
then yields an inhomogeneous first-order ODE determining the radial evolution of $\lt_1$. 
In particular, we can obtain the asymptotic radial behavior of $\lt_1$ by substituting in Eq. \eq{ltCorr}
the asymptotic expansions Eqs. \eqref{wtEta0} and \eqref{pbEta0}. We thereby obtain 
an inhomogeneous differential equation  of the form
\be
\lt'_1=-\frac{\lt_1}{r-1} -\frac{3}{r(r-1)}+ f_{l\,\cos}(r) + O(\eta)  \;, \label{Eql12}
\ee
where $f_{l\,\cos}(r)$ is a sum of terms involving oscillatory factors and powers of $r$.  
Solving the resulting equation yields the following expression for $\lt_1$
\bea
\lt_1(r)&=&\frac{A_{(l,\,\frac{1}{2}{\rm cos})}}{r^{1/2}}\cos{\left(  \frac{\sqrt{7}}{2}\ln{r} + \Phi^{l2} \right)} \nonumber \\
&&+ \frac{1}{r}\left[ A_{(l,\,1)} +A_{(l ,\,\ln)}\ln{r} \right. \nonumber \\
&& \left.+ A_{(l,\,{\rm cos})}\cos{\left( \sqrt{7}\ln{r} + \Phi^{l1} \right)} \right]   +O(\eta) \;. \nonumber \\ \label{llaw1}
\eea
This form of $\lt_1(r)$ is in agreement with the arguments of the beginning of this section (see \eqref{xinonlin}) concerning the form of perturbations around the fixed point \eqref{eigenval}. Indeed, one can see the oscillatory behavior which becomes dominant for large $r$.

However, the new information we got from the reasoning of the present section is that the unusual slowly-decaying,
oscillatory contributions enter the metric function $\lt$ (and therefore $L = L_S \lt$) {\it only at the  $O(\eta)$ level},
because $\lt = 1 + \eta \lt_1 + O(\eta^2)$. It is phenomenologically important to discuss also how these 
unusual slowly-decaying oscillatory contributions enter the other metric function, namely $g_{00} = - e^{2 \Phi}$,
or the related function $F= \Phi'$. As we showed above that $\lim_{\eta\to 0} F=F_S$, we immediately see that 
the unusual slowly-decaying oscillatory contributions will also  enter $F$, and thereby $\Phi$, only at the $O(\eta)$ level.

More precisely, considering the series expansion in $\eta$ of Eq. \eqref{Fsol} for $F$ one obtains an expression of the type
\be \label{Fexpeta}
F=F_S+\eta f_F(\lt_1, \wt_{\eta0}, \pb_{\eta0}) + O(\eta^2)\;,
\ee
where  $f_F$ is a  polynomial  function of its arguments. Substituting in this expression the  asymptotic behavior 
of the various arguments, namely Eqs. \eqref{wtEta0} and \eqref{pbEta0} for  $\wt_{\eta0}$ and $\pb_{\eta0}$,
and Eq.  \eqref{llaw1} for $\lt_1$,  one finds that the asymptotic radial behavior of $F(r)$ is of the form
\bea \label{Flaw2}
F(r)&=& F_S+\eta\left\{ \frac{A_{(F,\,\frac{1}{2}{\rm cos})}}{r^{3/2}}\cos{\left(  \frac{\sqrt{7}}{2}\ln{r} + \hat{\Phi}^{F2} \right)}  \right. \nonumber \\
&& + \frac{1}{r^2}\left[ A_{(F,\,1)} + A_{(F,\,\ln)}\ln{r}  \right. \nonumber \\
 && \left. \left. + \hat{A}_{(F,\,\cos)}\cos{\left(  \sqrt{7}\ln{r} + \hat{\Phi}^{F1} \right)} \right]  \right\}\nonumber \\
 &&+O(\eta^2) \;.
\eea
Having in hand such an explicit form of the asymptotic behavior of $F(r)$ one can now discuss the 
 physical consequences of having such slowly-decaying (and oscillating) contributions.
 More precisely, we can estimate the order of magnitude of the ``crossing distance", say $r_\times $, where  the oscillatory contributions in $F$
 [i.e., the terms $\propto \eta r^{-3/2} \cos \left(  \frac{\sqrt{7}}{2}\ln{r} + \hat{\Phi}^{F2} \right) $ in Eq. \eq{Flaw2}]
 start to dominate over the usual, Schwarzschildlike power-law decay contained in $F_S \sim r_h/(2 r^2)$.
 Using units where $r_h=1$, the coefficient of the oscillatory terms is $O(1)$ (because $\pb_0 = O(1)$).
 The crossing then happens when $F_S\sim 1/r_\times ^2 \sim \eta/r_\times^{3/2}$ which gives $r_\times  \sim \eta^{-2}$,
 i.e., coming back to general units
 \be
 r_\times  \sim \eta^{-2} r_h\,.
 \ee 
 Remembering the phenomenological constraint Eq. \eq{solarsystem} from solar-system tests, we see that the crossing
 scale $r_\times $ must be at least $10^{10}$ larger than the BH-horizon size. As, at such large distances, an astrophysical
 BH cannot be treated as being isolated, but rather embedded within a matter distribution created by many other astrophysical objects, 
 this suggests that the oscillatory contributions will not, in most cases, lead to identifiable observable signals. 
 This is phenomenologically fortunate for the compatibility of the torsion bigravity BH structure with observational data on
 astrophysical BHs (such as the existence and structure of accretion disks) because the dominance of the 
 $\propto \eta r^{-3/2} \cos \left(  \frac{\sqrt{7}}{2}\ln{r} + \hat{\Phi}^{F2} \right) $ in Eq. \eq{Flaw2} when $r \gtrsim r_{\times}$
 actually means that the gravitational force $F(r)= \Phi'(r)$ generated by the BH is no longer of the standard, attractive $F\simeq r_h/(2 r^2)$
 type, but starts oscillating around zero as: $ F\simeq r_h r_{\times}^{-1/2} r^{-3/2} \cos \left(  \frac{\sqrt{7}}{2}\ln{r} + \hat{\Phi}^{F2} \right) $. If $ r_{\times}$  was a distance scale where an accretion disk would be known to exist around the BH, this would
 phenomenologically rule out torsion-hairy BHs because the $\ln r$-periodic changes of sign of $F(r)$ when $r \gtrsim r_{\times}$ means 
 that the gravitational force created by the BH periodically becomes repulsive, before becoming again attractive, etc.
 
Let us complete our discussion of the asymptotic gravitational field of torsion-hairy BHs by commenting on
the presence of the logarithmic term $ F_{\ln} (r) =\frac{1}{r^2}A_{(F,\,\ln)}\ln{r} $ in $F$. If we use again units where 
where $r_h=1$, we will have $A_{(F,\,\ln)} = O(1)$. Therefore, the ratio between  $ F_{\ln} (r)$ and 
  the \sch term in \eqref{Flaw2} is of order
  \be
  \frac{F_{\ln} (r)}{F_S(r)} \sim \eta \ln \frac{r}{r_h}\,.
  \ee
  In view of the phenomenological limit \eq{solarsystem}, this ratio is very small, even if one considers 
  distances comparable to the crossing scale $r_\times \sim \eta^{-2} r_h$.

At face value, this preliminary discussion of the physical effects of the asymptotic behavior of $F(r)$, Eq. \eq{Flaw2},
seems to suggest that the solar-system upper limit \eq{solarsystem} constrains so much the magnitude of 
the unusual torsion-hairy gravitational field $F(r)$, that it will not lead to any  observable effect.
However, the more detailed discussion of the next section will show that such a conclusion would be premature.

\section{Phenomenological consequences of torsion bigravity for supermassive BHs}

We recalled in the Introduction some of the current observational 
data \cite{LIGOScientific:2019fpa,Akiyama:2019cqa,Abuter:2018drb,Abuter:2020dou} that probe (in a quantitative way) the structure
of the gravitational field of BHs. As our current investigation of torsion bigravity is limited to the static sector, we cannot
meaningfully discuss the predictions of torsion bigravity for coalescing BHs. [Such an investigation would anyway
require 3D numerical simulations to be conclusive.] 
Here, we focus on the phenomenological consequences of the hypothetical 
existence \footnote{We will, however, recall in the Conclusions that it is not clear at this stage
 whether gravitational collapse in torsion bigravity will generate the type of torsion-hairy BHs studied here,
 or, rather, ordinary GR BHs, which are also exact solutions of the torsion bigravity field equations.}
in our Universe of torsion-hairy asymptotically flat black holes. 
We successively consider the observable consequences of torsion-hairy BHs for : (i) the size of the shadow of BHs;
(ii) the redshift of the S2 star around supermassive black hole candidate Sgr A*; and (iii) the periastron
precession of S2.

\subsection{BH shadows}

The 2017 Event Horizon Telescope image of Messier 87 \cite{Akiyama:2019cqa} has inferred a size for the
shadow of the supermassive BH at the center of Messier 87 that agrees with GR predictions within 
$\sim 17 \%$ at the 68-percentile level. It has been emphasized in the recent work \cite{Psaltis:2020lvx}
that this measurement offers a new test of GR that goes beyond solar-system weak-field tests in probing
the strong-field structure of gravity. As discussed in Ref. \cite{Psaltis:2020lvx} the size of the BH shadow
can be discussed with sufficient accuracy within the context of non-rotating, spherically-symmetric BHs.
It then only depends on the time-time metric component $- g_{00}(r)= e^{2 \Phi(r)}$, considered as a function
of the areal radius $r$. The shadow radius $r_{\rm sh}$ is given by
\be \label{rsh}
r_{\rm sh} = r_{\rm lr} e^{- \Phi(r_{\rm lr})}\,,
\ee
where $ r_{\rm lr}$ denotes the {\it light ring} radius (or photon-sphere radius), which is the solution of the
equation
\be \label{rlr}
 r_{\rm lr} F( r_{\rm lr})=1 \,.
\ee
Here, as above, $F(r) = \Phi'(r)$ is the (relativistic) ``gravitational force" derived from the potential $\Phi(r)$.

In view of our general result Eq. \eq{Fexpeta} above,  we can a priori
see that the light ring radius of torsion-hairy BHs will admit an expansion of the form
\be
 r_{\rm lr}^{\rm TBG}=  r_{\rm lr}^{\rm GR}(r_h) (1 + a_1 \eta + O(\eta^2) )\,,
\ee
where $r_{\rm lr}^{\rm GR}(r_h) = \frac32 r_h$, and where the  coefficient $a_1$ is some $O(1)$ 
dimensionless function of the $O(1)$ shooting parameter $\pb_0$ 
describing the one-parameter family of torsion bigravity BHs having some given horizon radius $r_h$.
By the same reasoning, one also concludes (from Eq. \eq{rsh})
that
\be \label{rshTBG}
 r_{\rm sh}^{\rm TBG}=  r_{\rm sh}^{\rm GR}(r_h) (1 + b_1 \eta + O(\eta^2) )\,,
\ee
with another coefficient $b_1 = O(1)$.

Remembering now the strong solar-system limit, Eq. \eq{solarsystem}, on the magnitude of $\eta$, and comparing it to the
$\sim 17 \%$  precision on the Event Horizon Telescope estimate of $r_{\rm sh}$, we conclude that current (and foreseeable)
shadow measurements will not bring new constraints on the torsion bigravity theory parameter $\eta$.

As we are going to see next, the GRAVITY collaboration {\it weak-field} probing of the gravitational field of the Galactic central BH candidate
do provide (somewhat surprisingly) much stronger tests of torsion bigravity than the {\it strong-field} test provides by the 
Event Horizon Telescope Messier 87 observations.

\subsection{Redshift of S2}

An important observable object related to the supermassive black hole candidate Sgr A* is the star S2. Its highly elliptical (eccentricity 
$e=0.885$) orbit is characterized by a pericentre distance of about 1420 \sch radii, and an orbital speed at the pericentre point 
about $2.55\%$ of $c$.  This makes the star S2 a sensitive probe of gravity near a black hole. Two important quantitative results
based on observations of S2 star were recently obtained. The first one is a measurement of the gravitational redshift of S2 along its 
orbit \cite{Abuter:2018drb}. The second one is a measurement of the periastron precession of the orbit of S2 \cite{Abuter:2020dou}. 
In the present subsection, we shall study how the measurements of the gravitational redshift of S2 probe the gravitational
field structure around torsion-hairy BHs.

In the following we identify $r_h$ (in our torsion-hairy solutions) with the measured \sch radius $ r_h^{\rm obs}= 2 m_{BH} \equiv 2 G M_{BH}$
of the Galactic (candidate) BH. Here $M_{BH}$ denotes the mass of the black hole expressed in kg or $M_{\odot}$.

The reference paper we use is Ref. \cite{Abuter:2018drb}. In that paper, at any moment of time $t$, the redshift $z$ is decomposed in the following components,
\bea \label{redShFla}
z(t) \equiv \frac{\Delta \lam}{\lam} &=& B_0 - \frac{v_{\rm rad}(t)}{c} + f\left[ \frac{1}{2}\frac{v_{\rm orbit}^2(t)}{c^2} + \frac{m_{\rm BH}}{c^2 r(t)} \right]  \nonumber \\
&& -  \Phi_{\rm TBG}(t) + O(c^{-3}) \;. 
\eea
Here the radial velocity $v_{\rm rad}$ is the projection of the orbital velocity along the line of sight. The constant $B_0$ defines 
a constant offset. The second term (within square brackets) represents the classical GR redshift effect. It comprises
two contributions: the second-order Doppler effect, $\frac{1}{2}\frac{v_{\rm orbit}^2(t)}{c^2}$,  and and the Einstein gravitational redshift, $+ \frac{m_{\rm BH}}{c^2 r(t)} $.  The parameter $f$ was introduced in Ref. \cite{Abuter:2018drb} as a phenomenological
way to characterize the deviations  between Newtonian and relativistic physics: the value $f=0$ would
correspond to purely Newtonian physics, while the  value $f=1$ corresponds to GR. By generalizing the standard
derivation of the GR redshift \cite{Schrodinger:2010zz}, we have completed the GR result by adding, to lowest order,
the additional redshift $- \Phi_{\rm TBG}$ coming from torsion bigravity. It is defined, with our notation, as
\be \label{defPhiTBG}
 \Phi_{\rm TBG}(r) \equiv \Phi(r) - \Phi_S(r)\;.
\ee 
The observational results  presented in \cite{Abuter:2018drb} are summarized
in the following experimental constraint on the phenomenological parameter $f$:
\be \label{fredshift}
f=0.90\pm0.09|_{\rm stat}\pm0.15|_{\rm sys}. 
\ee
As the second-order Doppler effect, and the Einstein redshift, are fully degenerate, and have
the same amplitude of variation (because of energy conservation,  $\frac{1}{2}v_{\rm orbit}^2(t) - \frac{m_{\rm BH}}{ r(t)}= {\rm const.}$), 
the constraint \eq{fredshift} means any additional redshift correction from torsion bigravity would only have
been visible if it were larger that $2 \times \sqrt{0.09^2+0.15^2}= 0.34$ times the variation, along the orbit, of the
Einstein-Schwarzschild redshift $\Phi_S \approx - \frac{m_{\rm BH}}{ r(t)}= - \frac{r_h}{2 r(t)}$, namely
\bea \label{ratioredshift}
\Delta \Phi_S &\equiv&  |\Phi_S(r_{\rm ap}) - \Phi_S(r_{\rm per})| = \frac1{2840} -  \frac1{46550}\nonumber \\
&&= 3.306 \times 10^{-4}\,,
 \eea
 where $r_{\rm ap}= a (1+e)= r_{\rm per} \frac{1+e}{1-e} \approx 23275 r_h$ denotes the apocenter distance, and $r_{\rm per}= a (1-e)\approx 1420 r_h$ the pericenter distance.

As we have shown above that the deviations from Schwarzschild's metric in torsion-hairy BHs are proportional to $\eta$ (when
$\eta$ is small), we can consider, for concreteness, the case where $\eta=10^{-5}$. Choosing such a specific value (compatible
with solar-system tests (see \eq{solarsystem}), we numerically computed the ratio
\be \label{DefRatioPhi}
\frac{\Delta\Phi_{\rm TBG}^{\rm max}}{\Delta \Phi_S} \equiv \frac{ | \Phi_{\rm TBG}(r) -  \Phi_{\rm TBG}(r_{\rm ap})|^{\rm max}}{|\Phi_S(r_{\rm ap}) - \Phi_S(r_{\rm per})|}\,,
\ee
where the indicated maximization is done over $r$, as it ranges over the full orbit of S2.

After having fixed $\eta$ to the value $\eta=10^{-5}$, we still have a one-parameter family of asymptotically flat torsion-hairy BHs,
described by the horizon shooting parameter $\pb_0$. Not all values of $\pb_0$ lead to an asymptotically flat BH.
But, there is a rather large range of $\pb_0$, namely, the interval $-289<\pb_0< +5.5$ which are in the basin
of attraction of the relevant attractor Eq. \eq{Goldenfp}. We give in Table \ref{tab1} a sample of values
of the ratio $\frac{\Delta\Phi_{\rm TBG}^{\rm max}}{\Delta \Phi_S}$ defined in Eq. \eq{DefRatioPhi} as $\pb_0$
varies in this allowed interval. [When $\eta=10^{-8}$, the lower limit of the allowed interval is about $-267$.] We can summarize the results of Table \ref{tab1} in the inequality
\be \label{RatioRS}
0.004  \left(\frac{\eta}{10^{-5}}\right) \leq \frac{\Delta\Phi_{\rm TBG}^{\rm max}}{\Delta \Phi_S} \leq 0.047 \left(\frac{\eta}{10^{-5}}\right)\;. 
\ee
As said above, the current fractional sensitivity of redshift observations on S2 are of order of $0.34 \Delta \Phi_S$.
 Thus one can conclude that, near the supermassive black hole in SgrA*, the deviation of torsion bigravity from GR has, within the current observational precision, a negligible effect on redshift measurements.

\subsection{Periastron precession of S2}

Let us now discuss the modification of the periastron preseccion of S2 due to the difference between torsion bigravity and GR.
To compute the precession of the orbit of a star, we start from the geodesic equation [we recall that, in torsion bigravity, massive particles 
follow the geodesics of $g_{\mu \nu}$, rather than autoparallels of the torsionful connection ${A^{\lambda}}_{\mu \nu}$ \cite{Hayashi:1980av}.]
The geodesic equations of motion written in terms of coordinate time $t = x^0/c$ read
\be
\frac{d^2 x^i}{dt^2}=-\left( \Gamma^i_{\;\,\mu\nu} - \Gamma^0_{\;\, \mu\nu} \frac{dx^i}{c\,dt} \right)\frac{dx^\mu}{dt} \frac{dx^\nu}{dt}\;.
\ee
Expanding the right-hand side in the usual post-Newtonian way for the GR contributions, and working linearly in the
additional contributions due to torsion bigravity, we can rewrite the above equations of motion as
\be
\frac{d^2 x^i}{dt^2}= -\frac{ m_{\rm BH}x^i}{r^3} + a^i_{\rm GR} + a^i_{\rm TBG} + O(c^{-4})\;. \label{EqMotion}
\ee
Here the first term is the Newtonian gravitational force and would define (if it were alone) an elliptical Keplerian trajectory without any precession. The second term represents the first post-Newtonian correction (of order $c^{-2}$) to the Newtonian force coming from the GR
piece in the metric. [Here, we decompose the full metric as $g_{\mu \nu}= g_{\mu \nu}^S + h_{\mu \nu}^{\rm TBG}$,
and work linearly in the torsion bigravity metric correction $h_{\mu \nu}^{\rm TBG}$.]
The last term represents the corrections coming from the torsion bigravity. As usual in the post-Newtonian approach,
the leading-order additional contribution to the equations of motion only comes from the time-time component 
$c^2 h_{00}^{\rm TBG} \approx -  2\,c^2 \Phi_{\rm TBG}$, where $\Phi_{\rm TBG}$ was defined in Eq. \eq{defPhiTBG}.
This leads (at leading post-Newtonian order) to
\be \label{aTBG}
a^i_{\rm TBG} = -c^2 \partial_i \Phi_{\rm TBG}\;.
\ee
To calculate the periastron precession in the problem of motion defined by Eq. \eqref{EqMotion}, it is convenient to use
the standard Gauss form for the perturbation of Keplerian elements (see, e.g., Ref. \cite{CelMech}).
We are only interested in the secular advance of the argument $\omega$ of the periastron.
The Gauss form (see, e.g., Eq. (33) page 301 of \cite{CelMech}) yields $d \omega/dt$ as a linear expression
in the components of the perturbing accelerating force $a^i_{\rm GR} + a^i_{\rm TBG}$. 
To leading order, we have therefore 
\be
\frac{d \omega}{dt}= \left(\frac{d \omega}{dt}\right)^{\rm GR}+  \left(\frac{d \omega}{dt}\right)^{\rm TBG}\,,
\ee
where $\left(\frac{d \omega}{dt}\right)^{\rm GR}$ is the usual GR periastron precession, whose integral over one orbit
is the well-known value
\be \label{omGR}
\frac{\delta \omega^{\rm GR}}{2\pi} = \frac{3m_{\rm BH}}{c^2 a(1-e^2)}\;.
\ee
The expression of the additional contribution coming from torsion bigravity simplifies in the present case where 
the perturbing force $a^i_{\rm TBG}$ is purely radial, i.e., directed along the radial direction $n^i = \d_i r$. It reads \cite{CelMech}
\be
\left(\frac{d \omega}{dt}\right)^{\rm TBG}=-\frac{(1-e^2)^{1/2}}{n\, a\, e}\cos{\phi}\,\d_i r  a^i_{\rm TBG}  \label{DOmDt}
\ee
 Taking into account the conservation of angular momentum (per unit mass)
$$
r^2\frac{d \phi}{dt} = L = \sqrt{a(1-e^2)m_{\rm BH}}\,,
$$
where we used the expression for the angular momentum of the Newtonian elliptical motion,
and the expression of $a^i_{\rm TBG}$ in terms of the gradient of $\Phi_{\rm TBG}$, Eq. \eq{aTBG},
we get the following simple formula for the torsion-bigravity contribution to the 
 total (reduced) periastron precession, integrated over one full period,
 \bea \label{omTBG}
\frac{\delta \omega_{\rm TBG}}{2\pi} &=& \frac{1}{2\pi e m_{\rm BH}}\int_0^{2\pi}\cos{\phi}\,r^2\frac{\d \Phi_{\rm TBG}}{\d r}\,d\phi \nonumber \\
&&= \frac{1}{2\pi e m_{\rm BH}}\int_0^{2\pi}\cos{\phi}\,r^2F_{\rm TBG}(r)\,d\phi \;,\nonumber\\
\eea
where we replaced (as is allowed to leading order) $\Phi'_{\rm TBG}$ by 
\be
F_{\rm TBG}(r) \equiv F(r)-F_S(r).
\ee

We must now compare the value of  $\frac{\delta \omega_{\rm TBG}}{2\pi}$ to the usual GR periastron advance, Eq. \eq{omGR}.
More precisely, we should compare the torsion bigravity contribution $\frac{\delta \omega_{\rm TBG}}{2\pi}$ to the 
observational accuracy with which the periastron advance has been measured.

The GRAVITY Collaboration has recently succeeded to measure the integrated periastron precession of the S2 orbit \cite{Abuter:2020dou}. The result was presented in the form 
\be \label{omobs}
\delta \omega_{\rm obs}= f_{\rm SP}\frac{6\pi \, m_{\rm BH}}{c^2 a(1-e^2)} = 12.1' \,f_{\rm SP}\;.
\ee
where the coefficient $f_{\rm SP}$ was best-fitted  to the observations. The final result given 
in Ref. \cite{Abuter:2020dou} for the coefficient $f_{\rm SP}$  reads
\be
f_{\rm SP}=1.10 \pm 0.19\;.
\ee
In other words, the precision of the measurement is $19 \%$ of the GR value.

Now let us consider the contribution to periastron precession coming from the torsion bigravity corrections. 
In the case (discussed above) of  redshift observations, we found that torsion-hairy BHs induce  (when $\eta \lesssim 10^{-5}$)
only a negligible modification of the GR redshift. We might similarly expect, as all post-GR metric deviations around  torsion-hairy BHs
contain a factor $\eta$, so that  $F_{\rm TBG}(r) \equiv F(r)-F_S(r) = O(\eta)$ [see Eq. \eq{Fexpeta}],
that  $\delta \omega_{\rm TBG}$ will be automatically small compared to the error bar on $\delta \omega_{\rm obs}$, Eq. \eq{omobs}.
Surprisingly, we found that this is not the case.

Calculating the torsion-bigravity contribution by combining the explicit formula Eq. \eq{omTBG} 
with numerical simulations of torsion-hairy BHs (depending on the choice of shooting parameter $\pb_0$),
we found that, when $\eta=10^{-5}$, the ratio $\delta \omega_{\rm TBG}/\delta \omega_{GR}$ 
ranged (as $\pb_0$ varied) between $-21$ and $+160$, passing through $0$ for a certain value of $\pb_0$.
A sample of our numerical results is listed in Table \ref{tab1}.

\begin{widetext}

\begin{table}[h]
\caption{\label{tab1} Sample of redshift ratios and periastron ratios. The periastron ratios are computed  for two different values of $\eta$, namely $\eta=10^{-5}$ and $\eta=10^{-8}$.}
\begin{ruledtabular}
\begin{tabular}{||c c c c c c c c c||}
$\pb_0$ & $-250$ & $-220$ & $-180$ & $-160$ & $-100$ & $-20$ & $1$ & $5.5$ \\
\hline
$\Delta \Phi_{\rm TBG}^{\rm max}/\Delta\Phi_S|_{\eta=10^{-5}}$ & $0.047$ & $0.045$ & $0.039$ & $0.034$ & $0.024$ & $0.008$ & $0.004$ & $0.035$ \\
$\delta \omega_{\rm TBG}/\delta \omega_{GR}|_{\eta=10^{-5}}$ & $154.9$ & $85.5$ & $19.8$ & $1.5$ & $-21.2$ & $-15.0$ & $-6.5$ &
$1.8$ \\
$\delta \omega_{\rm TBG}/\delta \omega_{GR}|_{\eta=10^{-8}}$ & $0.168$ & $0.106$ & $0.025$ & $0.004$ & $-0.021$ &  $-0.015$ &  
$-0.006$ & $0.003$ \\
\end{tabular}
\end{ruledtabular}
\end{table}

\end{widetext}

Let us remark that, as one can see from the second line of this table, for  $\eta=10^{-5}$ and  certain values of $\pb_0$ , the periastron precession per orbit is enormous: $160 \times \omega_{GR} \approx 32^\circ $! Such a large deviation from GR is totally
incompatible with the recent result of the GRAVITY collaboration.

As already said, the contribution from torsion bigravity is (approximately) proportional to the value of $\eta$. Therefore, illustrated 
on the last row of Table \ref{tab1}, to pass the current tests of periastron precession one needs to take $\eta < 10^{-8}$. 
This shows that the periastron precession of S2 is a much more stringent test of the existence of torsion-hairy BHs than, both (i)
the other observations concerning supermassive BHs (redshift of S2 and shadow of Messier 87); and (ii) all solar-system tests.

The {\it a priori} surprising fact that the contribution to periastron precession can exceed one hundred times  the GR one, while the contribution to redshift is below the experimental precision,  has a simple technical explanation. 
Let us first recall that the gravitational redshift effect in GR is proportional to the Newtonian potential (see Eq. \eqref{redShFla}), 
while the value of the periastron precession in GR is proportional to the first post-Newtonian, $O(\frac1{c^2})$, correction to
the potential (see Eq. \eq{omGR}). By contrast, the torsion bigravity contribution to the 
potential gives additional contributions  {\it both} to the redshift (see Eq. \eq{redShFla}), and to periastron 
precession (see Eq. \eq{omTBG}). [A crucial fact being that  $F_{\rm TBG}(r) \approx \Phi'_{\rm TBG}(r)$ is {\it not} $\propto 1/r^2$,
and moreover decays more slowly than $1/r^2$, so that there is a large contribution to $\delta \omega_{\rm TBG}$ coming
from the apocenter of S2.] Then, it happens that the numerical magnitude of  $\Phi_{TBG}$ 
satisfies the following inequality 
\be \label{magnitudePhiTBG}
\left( \frac{m_{BH}}{c^2 r} \right)^2 \ll  \left(\frac{10^{-5}}{\eta} \right) \Phi_{TBG} \ll \frac{m_{BH}}{c^2 r}   \;.
\ee

\section{Conclusions}

We continued the study, initiated in Ref. \cite{Nikiforova:2020sac}, of spherically symmetric black hole solutions in torsion bigravity.
Ref. \cite{Nikiforova:2020sac} had shown that, in the infinite-range limit, these theories admit asymptotically flat black hole solutions  
because of the presence of attractive fixed points in the asymptotic (large-$r$) limit of the system of differential equations
describing the radial evolution of the metric and the torsion. We studied in detail the existence
and the structure of these fixed points. By considering the Jacobian matrix of the Fuchsian-type radial evolution
system near the fixed points, we discussed their attractive/repulsive character, as well as the way the metric
and torsion variables approach them at large radii. 

Several phenomenological aspects of asymptotically flat torsion-hairy black holes were
then considered: (i) location of the light ring and of the shadow; (ii) correction to the redshift of orbiting stars; 
and (iii) modification of the periastron precession of orbiting stars. Previous work \cite{Damour:2019oru}
had shown that, in the large-range limit, solar-system tests of relativistic gravity put the severe constraint
$\eta \lesssim 10^{-5}$ on the (single) theory parameter of  torsion bigravity. 
We showed that, when $\eta =10^{-5}$, the observable properties of asymptotically flat torsion-hairy black holes 
are automatically compatible with existing observational data on: (a) the shadow of Messier 87 \cite{Akiyama:2019cqa}; 
and (b) the redshift of the star S2 orbiting the Galactic-center massive black hole \cite{Abuter:2018drb}. See Eq. \eq{rshTBG}
and the first row of Table \ref{tab1}.

However, we found that when $\eta = 10^{-5}$, the one-parameter family of 
asymptotically flat torsion-hairy black holes (parametrized by the horizon parameter $\pb_0$)
generically induce an additional (post-GR) contribution to the periastron precession of S2 
which is typically much larger than the \sch precession. See second row of Table \ref{tab1}.
Barring a fine tuning of the value of the horizon shooting parameter $\pb_0$, we concluded that 
the current measurements of the periastron precession of S2 \cite{Abuter:2020dou}
are about a thousand times more stringent than  solar-system gravitational tests,
and constrain the torsion bigravity theory parameter $\eta$ to be (see third row of Table \ref{tab1})
\be \label{eta8}
\eta \lesssim 10^{-8}\,.
\ee
The physical reasons behind this surprisingly stringent probing power of periastron precession are briefly
discussed at the end of Section IV (see notably Eq. \eq{magnitudePhiTBG}).
 It is generally expected (given current solar-system tests
of weak-field gravity) that non-GR black holes will differ from GR ones mostly in the near-horizon,
strong-field regime \cite{Psaltis:2020lvx}. We note that torsion-hairy black holes
offer a counter-example to this expectation. Though, their observable predictions tend 
to the GR ones as $\eta \to 0$, the fact that the torsion-bigravity  gravitational field $F_{TBG}(r) \equiv F(r) - F_{\rm GR}(r)$
decays, at large radii, in a slower than $1/r^2$ manner 
(namely $\propto \eta r^{-3/2} \cos \left(  \frac{\sqrt{7}}{2}\ln{r} + \hat{\Phi}^{F2} \right) $, see Eq. \eq{Flaw2})
makes the periastron precession of S2 (whose orbit ranges between 1420 \sch radii and 23275 \sch radii)
a very sensitive probe of torsion-hairy black holes.

Let us also emphasize again that, contrary to what has been assumed in previous phenomenological discussions
of possible observable effects of spacetime torsion \cite{Mao:2006bb,March:2011ry}, the post-GR observable
effects of torsion-hairy black holes are not directly related to the torsion field around these black holes.
Our results in Section II allow one to derive the large-$r$ behavior of the two independent components
of the contorsion tensor, $K_{100}$ and $K_{122}$. 
For instance, we have  (using Eqs. (5.9) in Ref. \cite{Nikiforova:2020sac})
\be
K_{122}(r)= \left( \wt(r)-   \frac1{\lt(r)}  \right) W_S(r) , 
\ee
so that Eqs. \eq{wtEta0}, \eq{llaw1} above yield an asymptotic radial decay of $K_{122}(r)$ of the form
\be
K_{122}(r) = + \frac{2}{r} + O\left( \frac{\cos{\left( \frac{\sqrt{7}}{2}\ln{r}+\Phi^K \right)}}{r^{3/2}} \right).
\ee
However, this contorsion field, as well as the companion 
\be
K_{100}(r) = O\left( r^{-3/2} \cos \left( \frac{\sqrt{7}}{2}\ln{r}+\bar \Phi^K \right) \right)\,,
\ee
does not play a direct role in modifying the motion of orbiting stars. Indeed, test bodies in
torsion bigravity follow geodesics of the spacetime metric $g_{\mu \nu}$, and not autoparallels
of the torsionful connection ${A^{\lambda}}_{\mu \nu}$ \cite{Hayashi:1980av}.
[One would need to have, spin-polarized elementary fermions to directly probe the torsion hair.]
However, the presence of a propagating torsion field modifies the field equations, and thereby
indirectly modifies the structure of the metric $g_{\mu \nu}$. It is then the latter
torsion-induced metric modifications that have induced the slowly-decaying (and oscillating)
contributions to the gravitational force $F(r)=F_S(r) + F_{\rm TBG}(r)$ responsible for the
the unexpectedly large contributions to the periastron precession of S2.

As a final comment, let us emphasize that the very strong experimental constraint Eq. \eq{eta8} on the torsion-bigravity
theory parameter $\eta$
has been derived in the present work under the following two basic assumptions: (1) a vanishingly small inverse range $\k$;
and (2) real astrophysical black holes are described by the asymptotically flat torsion-hairy black hole solutions of
 torsion bigravity. The second assumption would be justified only if one proves that the collapse of a torsion-hairy
 star \cite{Damour:2019oru} in torsion bigravity does dynamically generate a torsion-hairy black hole, rather than
 an ordinary GR black hole (which is also an exact solution of torsion bigravity). It is quite possible (especially in view
 of the fact that \sch black holes  cannot support an infinitesimal torsion hair \cite{Nikiforova:2020sac}),
 that the torsion hair of a torsion-hairy star will be radiated away during the collapse. Another angle for 
 clarifying this issue is to study the dynamical stability of \sch black holes within the context of  torsion bigravity. 
 If, similarly to what was found in bimetric gravity \cite{Babichev:2013una,Brito:2013wya}, \sch black holes
 turn out to be unstable for small enough values of $\k r_h$, this will suggest that the torsion-hairy black holes
considered here will be the ultimate outcome of gravitational collapse. We leave this important issue to future work.

\section*{Acknowledgments}
I thank Thibault Damour  for informative discussions.

\appendix

\section{Explicit form of the field equations of static, spherically-symmetric torsion bigravity}\label{appA}

\bea
F&=&\left\{   3 - 3 r + 2 ( r-1) \eta  \lt  [-6 + (1 + \eta ) \pb ] \wt  \right. \nonumber \\
&&   - 
   \lt ^2 \left[\textcolor{white}{\lt} 2 r \eta  (1 + \eta ) \pb  + 
      k^2 r^3 \eta  (1 + \eta ) \pb ^2  \right. \nonumber \\
      && \left. \left. - 
      3 (r (1 + \eta ) + 3 ( r-1) \eta  \wt ^2)\textcolor{white}{\lt}\right]\right\}/\nonumber \\
      && \left\{ 2 ( 
     r-1) r [3 + \eta  \lt  (3 + (1 + \eta ) \pb ) \wt ]\right\} \label{Fsol}
\eea

\bea \label{plwsystem}
 \frac{d}{dr}\pb(\lt,\wt, \pb)  &=& \frac{3 - 3 \lt \wt - (1 + \eta) \pb (2 + \lt \wt)}{r (1 + \eta)} \;, \label{Dpb}\\
 \frac{d}{dr}\lt(\lt,\wt, \pb)  &=& \frac{N_{\lt}}{D_{\lt}} \;, \label{Dlt} \\
 \frac{d}{dr}\wt(\lt,\wt, \pb)  &=& \frac{N_{\wt}}{D_{\wt}} \;, \label{Dwt}
\eea
where
\begin{widetext}
\bea
N_{\lt}=&&  \lt  \left\{ 9 [\,9 (r + 4  \eta  - 3 r  \eta ) + 
      12 (r-1)  \eta  (1 +  \eta ) \pb + (r-2)  \eta  (1 +  \eta )^2  \pb ^2\,] - 
   3  \eta  \lt [\, 27 (5 + 6 r ( \eta-1 ) - 7  \eta ) \right. \nonumber \\
   &&- 
      9 (1 +  \eta ) (1 + (-3 + 4 r)  \eta )  \pb  - 
      3 (-9 + 8 r)  \eta  (1 +  \eta )^2 \pb^2 + (-1 + 2 r)  \eta  (1 +  \eta )^3  \pb ^3 \,]  \wt  \nonumber \\
      && + 
   2  \eta   \lt ^{\,3} (3 + (1 +  \eta )  \pb ) \wt [\, \k^2 r^3  \eta  (1 +  \eta )^2 (-1 + 2  \eta )  \pb ^3 + 
      3  \eta  (1 +  \eta )^2 \pb^2 (\k^2 r^3 + 2 r  \eta  - 2 ( r-1)  \eta   \wt ^2) \nonumber \\
      &&+ 
      27 (-r (1 +  \eta )^2 + ( r-1) ( \eta-1 )  \eta  \wt^2) - 
      9  \eta  (1 +  \eta ) \pb (r (-1 + \k^2 r^2 -  \eta ) + ( r-1) (1 +  \eta ) \wt^2) \,] \nonumber \\
      &&+ 
   3  \lt ^{\,2} [\, 2 r  \eta  (1 +  \eta )^2 ( \eta  (1 +  \eta ) + 
         \k^2 r^2 (-1 + 2  \eta ))  \pb ^3 + 
      \k^2 r^3  \eta ^2 (1 +  \eta )^3  \pb ^4 \nonumber \\
      && + 
      18  \eta  (1 +  \eta ) \pb (-\k^2 r^3 + ( r-1) (\eta-1 )  \wt ^2) - 
      3  \eta  (1 +  \eta )^2 \pb^2 (r + \k^2 r^3 - 3 r  \eta  + 
         11 ( r-1)  \eta   \wt ^2) \nonumber \\
         &&  \left.  + 
      27 (-r (1 +  \eta )^2 + (r-1)  \eta  (1 + 13  \eta ) \wt^2) \,] \textcolor{white}{\lt} \right\}  \;,\nonumber \\
D_{\lt}=&& 6 (1-r) r (1 + \eta) (-9 + \eta (1 + \eta) \pb^2) [3 + \eta \lt (3 + (1 + \eta) \pb) \wt] \;,
\nonumber \\
N_{\wt}=&& \k^2 r^3  \eta ^2 (1 +  \eta )^4  \lt ^{\,3}  \pb ^5  \wt  - 
 9 (1 +  \eta )^2 \pb^2 [\, 2 - 
    2 r + (8 - 9 r)  \eta   \lt   \wt  +  \eta   \lt ^{\,3}  \wt  (r + 3 \k^2 r^3 - 3 r  \eta  + 3 (-1 + r)  \eta   \wt ^2) \nonumber \\
    &&+ 
     \lt ^{\,2} (2 r (1 + 2 \k^2 r^2 -  \eta ) + (-9 + 10 r)  \eta  \wt^2) \,] + 
 81 [\, 4 - 4 r + (-6 + 5 r + 6  \eta  - 7 r  \eta )  \lt   \wt  -  \eta  (13 - 12 r +  \eta )  \lt ^{\,2}  \wt ^2 \nonumber \\
 && + 
     \lt ^{\,3}  \wt  (-r (1 +  \eta )^2 + (-1 + r) (-3 +  \eta )  \eta  \wt^2) \,] +  \eta  (1 +  \eta )^3  \lt ^{\,2} \pb^4 \left\{  6 \k^2 r^3 +  \eta  (1 +  \eta )  \wt ^2 + 
     \lt   \wt  [\, r (2  \eta  (1 +  \eta ) \right. \nonumber \\
     && \left. + \k^2 r^2 (-2 + 7  \eta )) - 
       2 (-1 + r)  \eta  (1 +  \eta )  \wt ^2\,] \textcolor{white}{\lt} \right\} + 
 3 (1 +  \eta )^2  \lt  \pb^3 \left\{ \textcolor{white}{\lt} -(-2 + r)  \eta  (1 +  \eta )  \wt   \right.  \nonumber \\
 &&+  \eta   \lt ^{\,2}  \wt  (r (-1 + 3 \k^2 r^2 (-1 +  \eta ) + 4  \eta  + 5  \eta ^2) - 
       5 (-1 + r)  \eta  (1 +  \eta )  \wt ^2) + 
    2  \lt  [\, r (2  \eta  (1 +  \eta ) + 
          \k^2 r^2 (-1 + 2  \eta )) \nonumber \\
          && \left. - (-1 + 
          r -  \eta )  \eta  (1 +  \eta )  \wt ^2\,] \textcolor{white}{\lt} \right\} - 
 27 (1 +  \eta ) \pb [\, 4 - 4 r + (r + 4  \eta  - 3 r  \eta )  \lt   \wt  + 
     \lt ^{\,3}  \wt  (r (1 + 2  \eta  + 2 \k^2 r^2  \eta  +  \eta ^2) \nonumber \\
     && - (-1 + 
          r) (-3 +  \eta )  \eta   \wt ^2) + 
    2  \lt ^{\,2} (r (1 + \k^2 r^2 +  \eta ) +  \eta  (3 - 
          2 r +  \eta )  \wt ^2) \,]  \;, \nonumber \\
D_{\wt}=&& 2 (1 - r) r (1 +  \eta )  \lt  (3 + (1 +  \eta )  \pb ) (-9 +  \eta  (1 +  \eta )  \pb ^2) [\, 3 +  \eta   \lt  (3 + (1 +  \eta )  \pb )  \wt \,] \;.
\eea
\end{widetext}


\end{document}